\documentclass[epj,referee]{svjour}
\usepackage{graphics}
\usepackage{bm}

\newcommand{\x}{{\bm x}}

\newcommand{\be}{\begin{equation}}
\newcommand{\ee}{\end{equation}}
\begin{document}
\title{Capillary filling with pseudo-potential binary Lattice-Boltzmann model}

\author{S. Chibbaro \inst{1,2} }
\offprints{chibbaro@iac.rm.cnr.it}          

\institute{                    
  \inst{1} Istituto per le Applicazioni del
  Calcolo CNR, Viale del Policlinico 137, 00161 Roma.\\
  \inst{2} Dept. of Mechanical engineering, University of Tor Vergata, 
 00133 Roma, Italy.
}
\date{Received: date / Revised version: date}
%

\abstract{
We present a systematic study of capillary filling for a
  binary fluid by using a mesoscopic  lattice Boltzmann model
for immiscible fluids
  describing a diffusive
  interface moving at a given contact angle with
  respect to the walls. The phenomenological way to impose a given contact angle is 
analysed. Particular attention is given to the case of complete wetting, that is contact angle equal to zero. Numerical results yield quantitative agreement with the
  theoretical Washburn's law, provided that the correct ratio of the dynamic viscosities 
between the two fluids is used. Finally, the presence of precursor films is experienced and it is shown that these films advance in time with a square-root law but with a different prefactor with respect to the bulk interface.
}
\authorrunning
\titlerunning
\maketitle

\section{Introduction}
Capillary filling is an important subject of research
for its  relevance to  microphysics and nanophysics
\cite{DeG_84,tas}, which deal with the development of modern nanofluidic devises, as well as with various applications of porous materials.
The fundamental physics of capillary filling has been widely studied from
the pioneering works of Washburn \cite{Wash_21} and Lucas
\cite{Lucas} and  the basic equations of such problem have been put forward
since decades \cite{Bos_23,Ske_71}.
However, since capillary filling is a
typical ``contact line'' problem, where subtle 
non-hydrodynamic effects take place at the contact point between liquid-gas
and solid phase, it remains an important subject of experimental and theoretical present research at micro and nano-scales \cite{Mas_02,Sup_03,Yan_04,Jua_06}.
Usually, only the late
asymptotic stage is studied, leading to the well-known Washburn's 
law, which predicts the following relation for the position of the 
interface inside the capillary:
\begin{equation}
z^2(t) - z^2(0) = \frac{\gamma H cos(\theta)}{3 \mu }   t 
\label{eq:wash}
\end{equation}
where $\gamma$ is the surface tension between liquid and gas, $\theta$
is the {\it static} contact angle, $\mu$ is the liquid dynamic viscosity, $H$
is the channel height and the factor $3$ depends on the geometry of
the channel (here a two dimensional geometry given by two infinite
parallel plates separated by a distance $H$ -- see fig. \ref{fig.1}). 

Experiments of such phenomena are quite difficult, mainly at nanoscopic scales, and thus 
numerical simulations are an essential tool for the investigation
of this problem. 
One possible approach is based upon an atomistic description,
where Newton's equations are integrated for a set of molecules,
which interact typically via a Lennard-Jones potential~\cite{frenkel}. 
This idea is behind the molecular dynamics (MD) approach,
which is regarded to reproduce rather well the reality. 
However, its computational efficiency is very low and macroscopic scales are 
very hard to reach.
Therefore, different models have to be investigated. 
The Lattice-Boltzmann (LB) method~\cite{Ben_92,Che_98,wolf} has been materialising during the last decades
as an efficient and powerful mesoscopic way to simulate complex flows.
Recently, this approach has been also assessed for the description 
of fluid-solid interactions, which have a major role for micro- and nanoscopic applications~\cite{Squ_05,Tab_03}.
In particular, different models for slip boundary conditions and wetting properties
have been put forward~\cite{Kan_02,Dup_05,Sbr_06,Har_06,Fab_07}.
Given that the LB phenomenological description is not based on atomistic details
but on average properties, this mesoscopic model looks promising
for those applications which can not be described within a pure continuum framework
but do not require a fully atomistic treatment.
Many interesting and challenging phenomena can be placed in this category,
like impact of a droplet on solid walls, capillary dynamics and water repellency on structured surfaces~\cite{Yar_06,Rey_06,Ric_02}. 

In this work, we propose to use the Lattice-Boltzmann model  (LBM) with a pseudo-potential interaction, 
first proposed by Shan and Chen \cite{SC_93,SS_92} 
and already successfully used in many applications in micro/nanoscience~\cite{Sbr_06,Kwok_05,Kwok_06}.
Specifically, we use the model proposed for the description 
of immiscible multi-fluids, limiting ourselves to a binary mixture.
For this kind of model, some numerical simulations have been worked out,
although in simplified configurations \cite{Wol_04}.
Nevertheless, this kind of models seems to assure a good compromise between 
efficiency and accuracy.
On the one hand, these models are hydrodynamic and, therefore, assure the correct 
macroscopic model, since there are many evidences that the capillary imbibition
remains hydrodynamic down to nanoscales \cite{Dim_07,Hub_07}.
On the other hand, the formulation in terms of a discretised Boltzmann equation
allows a high efficient and easily parallel algorithm  and
the possibility of introducing in a direct manner some ingredients which are taken 
from more microscopic approaches.
The study of capillary imbibition was recently attempted via Shan-Chen one-component 
(multi-phase)
LB method~\cite{Fab_07},
which has been found to reproduce correctly the basic physics of the capillary filling.
Nevertheless, this was found to be true only provided that the evaporation-condensation effect is negligible,
that is when the gas phase density is negligible with respect to the liquid one~\cite{Fab_07}.
In this sense, the binary  LBM seems to present a more general framework 
with respect to the multi-phase one, since the evaporation-condensation effect is 
very low for this model.

In this paper, we show the first evidence that the LBM chosen can satisfactorily reproduce 
 the physics of the capillary filling in realistic configurations (for example Water-air mixtures) with a proper choice of the physical parameters.
A given contact angle can be easily imposed through the introduction of 
phenomenological force between fluids and walls. The interesting situation of a complete 
wetting is also supported by this approach. The expected Washburn's law is found for the displacement of the interface between the two fluids. 
Moreover, precursors films appear to be present and to follow a Washburn's law, with a different pre-factor. 
All the methodological features  are detailed.

\section{Model}
In this work we use the multicomponent LB model proposed by Shan and Chen \cite{SC_93}. 
This model allows for  distribution functions of an arbitrary number of components, with different molecular mass:
\begin{eqnarray}
\label{eq:lbe}
f^k_i(\x+ {\bm c}_{i} \Delta t,t+\Delta t)&-&f^k_i(\x,t)= \\ \nonumber
&-&\frac{
  \Delta t}{\tau_k} \left[f^k_i(\x,t)
  -f_i^{k(eq)}(\x,t)\right] 
\end{eqnarray}
where $f^k_i(\bm{x},t)$ is the kinetic probability density function
associated with a mesoscopic velocity $\bm{c}_{i}$ for the $k$th fluid,
 $\tau_k$ is a mean
collision time of the $k$th component
(with $\Delta t$ a time step), and $f^{k(eq)}_{i}(\x,t)$
the corresponding equilibrium function.
The collision-time is related to kinematic viscosity by the formula
$\nu_k=\frac{1}{3}(\tau_k-\frac{1}{2})$.
For a two-dimensional 9-speed LB model (D2Q9)
$f^{k(eq)}_{i}(\x,t)$ takes the following form \cite{wolf}:
\begin{eqnarray}
\label{eq:feq1}
f^{k(eq)}_{0}&=&\alpha_kn_k-\frac{2}{3}n_k{\bf u}_k^{eq}\cdot{\bf u}_k^{eq}   \\
f^{k(eq)}_{i}&=&\frac{(1-\alpha_k)n_k}{5}+\frac{1}{3}n_k{\bf c}_i\cdot{\bf u}_k^{eq}   \label{eq:feq2} \\
&+& \frac{1}{2}n_k({\bf c}_i\cdot{\bf u}_k^{eq})^2-\frac{1}{6}n_k{\bf u}_k^{eq}\cdot{\bf u}_k^{eq} \;\;\;\textrm{for i=1$\ldots$4}  \nonumber \\
f^{k(eq)}_{i}&=&\frac{(1-\alpha_k)n_k}{20}+\frac{1}{12}n_k{\bf c}_i\cdot{\bf u}_k^{eq}    \label{eq:feq3} \\
&+& \frac{1}{8}n_k({\bf c}_i\cdot{\bf u}_k^{eq})^2-\frac{1}{24}n_k{\bf u}_k^{eq}\cdot{\bf u}_k^{eq} \;\;\;\textrm{for i=5$\ldots$8} \nonumber 
\end{eqnarray}
In the above equations ${\bf c}_i$'s are discrete velocities, defined as follows 
\be
{\bf c}_i= \left \{ \begin{array} {l} 0,  i=0, \\
\left(cos\frac{(i-1)\pi}{2},sin\frac{(i-1)\pi}{2}\right), i=1-4 \\
\sqrt{2}\left(cos[\frac{(i-5)\pi}{2}+\frac{\pi}{4}],sin[\frac{(i-5)\pi}{2}+\frac{\pi}{4}]\right),  i=5-8 
\end{array}
\right.
\ee
in the above, $\alpha_k$ is a free parameter related to the sound speed of the $k$th component, according to $(c_s^k)^2=\frac{3}{5}(1-\alpha_k)$; $n_k=\sum_if^k_i$ is the number density of the $k$th component. The mass density is defined as $\rho_k=m_kn_k$, and the fluid velocity of the $k$th fluid ${\bf u}_k$ is defined through $\rho_k{\bf u}_k=m_k\sum_i{\bf c}_if_i^k$, where $m_k$ is the molecular mass
of the $k$th component.
The equilibrium velocity  ${\bf u}_k^{eq}$ is determined by the relation
\be
\rho_k{\bf u}_k^{eq}=\rho_k {\bf U} +\tau_k{\bf F}_k
\ee
where ${\bf U}$ is the common velocity of the two components.
To conserve momentum at each collision in the absence of interaction (i.e. in the case of ${\bf F}_k=0$)
${\bf U}$ has to satisfy the relation 
\be
{\bf U}=\left(\sum_i^s \frac{\rho_k{\bf u}_k}{\tau_k}\right)/
\left(\sum_i^s \frac{\rho_k}{\tau_k}\right)\;.
\ee
The interaction force between particles is the sum of
a bulk and a wall components.
The bulk force is given by
\be
\label{forcing} 
{\bf F}_{1k}({\bf x})=-\Psi_k(\x)\sum_{\x^{\prime}} \sum_{\bar{k}=1}^sG_{k \bar{k}}\Psi_{\bar{k}} (\x^{\prime})(\x^{\prime}-\x) 
\ee
where $G_{k\bar{k}}$ is symmetric and $\Psi_k$ is a function of $n_k$.
In our model, the interaction-matrix is given by
\be
G_{k \bar{k}}=\left \{ \begin{array} {l} g_{k \bar{k}}, |\x^{\prime}-\x|=1, \\
g_{k \bar{k}}/4, |\x^{\prime}-\x|=\sqrt{2}, \\
0, \textrm{otherwise}. \end{array}
\right.
\ee
where $g_{k \bar{k}}$ is the strength of the interparticle potential between 
components $k$ and $\bar{k}$. In this study, the effective number density $\Psi_k(n_k)$ is taken simply
as $\Psi_k(n_k)=n_k$. Other choices would lead to a different equation of state (see below).

At the fluid/solid interface, the wall is regarded as a phase with constant number density.
The interaction force between the fluid and wall is described as 
\be
\label{forcingw} 
{\bf F}_{2k}({\bf x})=-n_k(\x)\sum_{\x^{\prime}} g_{k w} \rho_{w}
 (\x^{\prime})(\x^{\prime}-\x)
\ee
where $\rho_w$ is the number density of the wall and $g_{kw}$ is the interaction strength between component $k$ and the wall. By adjusting $g_{kw}$ and  $\rho_w$ , different wettabilities can be obtained. 
This approach allows the definition of a static contact angle $\theta$, by introducing   a
suitable value for the wall density $\rho_w$ \cite{Kan_02},
which can span the range $\theta \in [0^o:180^o]$.
It is worth noting that, with this method, it is not possible to know ``a priori'' the value of the contact angle 
from the phenomenological parameters.
Thus, an ``a posteriori'' map of the value of the static contact angle versus the value of the interaction strength $g_w$ 
has to be obtained.
To this aim, we have carried out several 
simulations of a static droplet attached to a wall
for different values of $g_w$~\cite{Kan_02,ben_06}.
In particular, in our work, the value of the static contact angle has been 
computed directly 
as the slope of the contours of near-wall density field, and independently
through the Laplace's law,
$\Delta P= \frac{2 \gamma cos \theta}{H}$.
The value so obtained is computed within an error  $\sim 2\%-3\%$.
Recently, a different approach has been proposed, which is able to give an ``a priori'' estimate
of the static contact angle from the phenomenological parameter~\cite{Hua_07}.
Nevertheless, we have preferred to retain our ``a posteriori'' method for its simplicity and efficiency.

In a region of pure $k$th component, the pressure
is given by $p_k=(c_s^k)^2m_kn_k$, where $(c_s^k)^2=\frac{3}{5}(1-\alpha_k)$.
To simulate a multiple component fluid with different  densities,
we let $(c_s^k)^2m_k=c_0^2$, where $c_0^2=1/3$.
Then, the pressure of the whole fluid is given by 
$p=c_0^2\sum_kn_k+\frac{3}{2}\sum_{k,\bar{k}}g_{k,\bar{k}}\Psi_k\Psi_{\bar{k}}$, which represents a non-ideal gas law.

The Chapman-Enskog expansion \cite{wolf} shows that the fluid mixture 
follows the Navier-Stokes equations for a single fluid:
\begin{eqnarray}
\partial_{t}\rho + \nabla \cdot (\rho {\bf u}) &=&0, \\
\rho [ \partial_{t}
    {\bm u} + ({\bm u} \cdot {\bm \nabla} ){\bm u}] &=& - {\bf \nabla}
  {P} +{\bm F} + {\bm \nabla} \cdot (\mu(
                     {\bm \nabla} {\bm u}+{\bm u}{\bm \nabla}) \nonumber
\label{eq:NS}
\end{eqnarray}
where $\rho=\sum_k \rho_k$ is the total density of the fluid mixture, 
the whole fluid velocity ${\bf u}$ is defined by
$\rho{\bf u}= \sum_k \rho_k{\bf u}_k+\frac{1}{2}\sum_k{\bf F}_k$
and the dynamic viscosity is given by $\mu= \rho \nu= \sum_k\mu_k = \sum_k(\rho_k\nu_k)$.

\subsection{Theoretical analysis}
For the geometry set up considered in this work, see fig. \ref{fig.1},
we propose a general model which can be integrated
 (in this paper we have used the software ``Mathematica'')
to provide a semi-analytical solution to be compared with numerical results.

The Washburn's law
(\ref{eq:wash}) holds when inertial forces can be neglected
with respect to the viscous and capillary ones \cite{Bos_23}.  
This is not true in the early stage of the filling process, 
typically a few nanoseconds for micro-devices, where strong
acceleration drives the interface inside the capillary.  
Another important effect  is the  ``resistance'' of the gas
occupying the capillary during the liquid invasion.
This resistance is due to the finite value of the gas density
and is a sensitive issue for numerical simulations, because reaching
the typical $1:1000$ density ratio between liquid and gas of experimental 
set up, represents a challenge for most numerical methods. 
In order to take into account both effects, inertia and gas dynamics,
one may write down the balance between the total momentum change
inside the capillary and the force (per unit width) acting on the 
liquid$+$gas system:
\begin{equation}
\frac{d (\dot z M(t))}{dt} = F_{cap}+F_{vis}
\label{eq:momentum}
\end{equation}
where  $M(t)=M_g+M_l$ is
the total mass of liquid and gas inside the capillary at any given
time. The two forces in the right hand side correspond to the capillary force,
$F_{cap} = 2 \gamma cos(\theta)$, and to the viscous drag $F_{vis} =
-2 (\mu_g(L-z) + \mu_lz)\partial_nu(0)$. 
We consider the Poiseuille solution for a channel, namely:
$$
u(y) = 6 \frac{ \bar{u} }{H^2} y(H-y)
$$
where $\bar u = \frac{d l}{dt} = 1/H \int dy u(y) $ is
the interface mean velocity,
\be
\partial_y u|_{y=0} = 6 \frac{\bar u}{H }
\ee
Inserting this expression in the viscous force and following the notation of
fig.(\ref{fig.1}), one obtains the final expression (see also
\cite{napoli,Fab_07} for a similar derivation:
\begin{eqnarray}
\label{GWE}
&&(\rho_g(L-z)+\rho_l z)
\ddot z  +  (\rho_l-\rho_g) (\dot z)^2  =  \nonumber \\
 &&2
  \frac{\gamma cos(\theta)}{H} - \frac{12 \dot z}{H^2} [(\mu_g (L-z) + \mu_l z)]\label{eq:front1}
\end{eqnarray}
In the above equation for the front dynamics, the terms in the LHS
take into account for the fluid inertia. 
Since these terms are  proportional either to the
acceleration or to the squared velocity, they become negligible for
long times, as the velocity becomes negligible too.
The right hand side takes into account for the forces.
The presence of two viscous forces related to both liquid and gas components
indicates that a gas resistance is felt for $\mu_g/\mu_l \ne 0$.
Thus, Washburn's law (\ref{eq:wash}) is correctly recovered 
asymptotically, for $t \rightarrow \infty$ and in the limit $\mu_g/\mu_l \rightarrow 0$.

\section{Numerical Results}
\begin{figure}
\resizebox{1.0\columnwidth}{!}{%
  \includegraphics{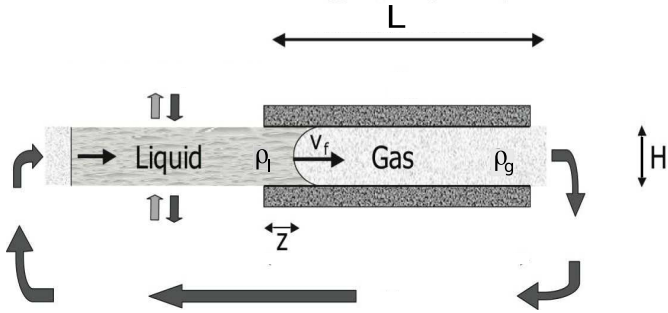}
}
\caption{Geometrical set-up of the numerical LBE. The two dimensional
     geometry, with length $2L$ and width $H$, is divided in two
     parts. The left part has top and bottom periodic boundary
     conditions such as to support a perfectly flat gas-liquid
     interface, mimicking a ``infinite reservoir''. In the right
     half, of length $L$, there is the true capillary:  the top and
     bottom boundary conditions are those of a solid wall, with a
     given contact angle $\theta$. Periodic boundary conditions are also imposed at the west and east sides. The figure is based on a figure in~\cite{Fab_07}.}
\label{fig.1}
\end{figure}

In figure \ref{fig.1} the set-up of our numerical experiments is sketched.
In order to simulate a realistic capillary we impose that
the bottom and top surfaces are coated only in the
right half of the channel with a boundary condition imposing a given
static contact angle \cite{Kan_02}; in the left half we impose
periodic boundary conditions at top and bottom surfaces in order to
 mimic an ``infinite
reservoir''. Periodic boundary conditions are also imposed at the two
lateral sides such as to ensure total mass conservation  inside the
system. 
At the solid surface, bounce back boundary conditions for the particle distributions were imposed.
It is important to note that, in the following, we shall indicate the first component with an index $1$, 
and the second with $2$, for the sake of clarity, while in figure \ref{fig.1}
they are respectively indicated by l (liquid) and by g (gas). 
The dimension of the channel have been chosen to be:
$L=200 \Delta z~;~~H=40 \Delta z$.
It is important to note that in LB methods the interface $\delta \xi$ is diffuse ($\delta \xi \approx 4-5$ grid points in present study) and it has been pointed out that meaningful results can be obtained only when the ratio $\delta \xi/H$ is large enough, $\delta \xi/H \sim 10$~\cite{Fab_07}.
Therefore, the size of the channel has been chosen such as to fulfill this condition.

The first issue to address in order to have a realistic simulation of a capillary filling, is 
the correct description of the physical properties of the system.
It is well known that in LB methods it is difficult to maintain realistic density ratio between different species.
 This is particularly true for binary LB methods, where almost all previous simulations have been limited to density ratio equal to 1 \cite{SC_93,Kan_02,Rot_07,Yeo_03}.
As a first validation, we present the results obtained for a dynamic configuration
where fluid and gas have equal density ($\rho_1=\rho_2=1$) and 
equal viscosities ($\mu_1=\mu_2=\mu=\frac{1}{6}$). 
The contact angle is chosen to be $\theta\approx 40$, that is  $g_{1w}=0;~ g_{2w}=0.2; ~\rho_{w}=0.7$.
This is the case so far considered in literature \cite{Kan_02,Wol_04}. 
For this case, by taking $\theta$ constant in time, 
a simple analytical solution of equation (\ref{eq:front1}) can be obtained:
\be
z(t)= \frac{\gamma H cos \theta }{6 \mu L}t_{d}\left[ \exp(-t/t_{d}) 
+ t/t_{d} - 1 \right] +z_0,
\label{eq:analy}
\ee
where $z_0$ is the starting point of the interface at the beginning of the simulation, 
$t_{d}=\frac{\rho H^2}{12\mu}$ is a typical transient time.
The analytical solution shows an asymptotic linear dependence in time for the front displacement $z(t)$. 
This behaviour is correctly reproduced 
by our LB model, as shown in figure \ref{fig.3}, where the numerical solution is compared
against the analytical solution given by eq. (\ref{eq:analy}).

\begin{figure}
\resizebox{1.0\columnwidth}{!}{%
\includegraphics{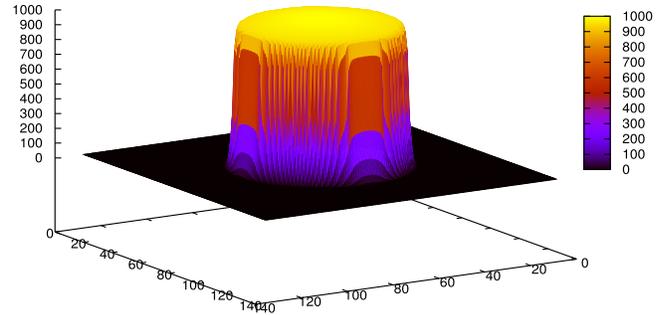}
}
\caption{Single-droplet total density profile. The density ratio between the droplet and the surrounding fluid is 1000, while the interface remains limited to a few grid points.}
\label{fig.2}
\end{figure}

\begin{figure}
\resizebox{1.0\columnwidth}{!}{%
  \includegraphics{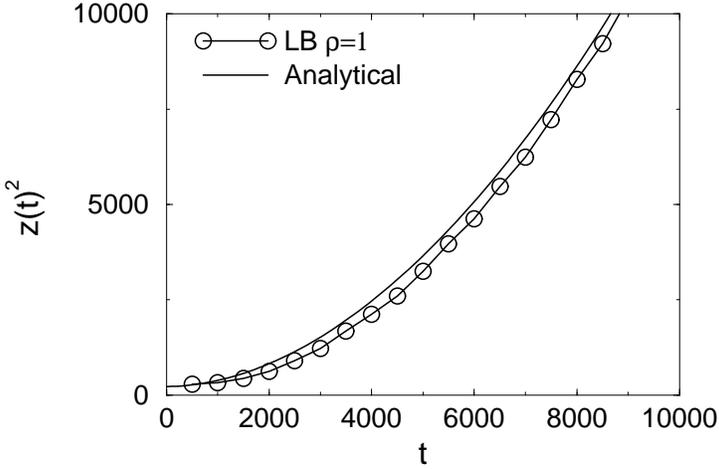}
}
\caption{The dashed line represents the numerical results for the case of two fluids with equal density and viscosity. The circled line represents the numerical simulation, while the solid line shows the analytical solution (\ref{eq:analy}) for the same physical parameters of the LB simulation. 
As expected, in this case, the front advances linearly in time and  $z^2(t)$ is a parabola.
Remarkably, LB results reproduces rather well also the initial transient in time, which is governed by an exponential term, see eq. (\ref{eq:analy})}
\label{fig.3}
\end{figure}

Recently, the issue of simulating components of different densities
has been addressed for problems of binary diffusion \cite{Kar_06}.
However, in our knowledge, there is no complete study of a dynamical phenomenon 
with components with different densities. 
Following a recent work \cite{McC_04},
it is pointed out that multicomponent LBE approach describes
two isothermal perfect gases, in mutual interaction, which conserve their 
isothermal thermodynamics.
In an isothermal gas, the sound speed is inversely proportional
to the molecular mass $c_s^2=\frac{\Gamma R T}{m}$, where $\Gamma$ is the gas adiabatic constant
 and $R$ the universal gas constant. 
In our model, molecular mass can be changed for each component 
together with the value of the parameter $\alpha_k$ in order to change  the value
of the sound speed in eqs. (\ref{eq:feq1})-(\ref{eq:feq3}).
For a binary mixture,
the two sound speeds are related as
$c_s^1=\sqrt{\frac{m_2}{m_1}}c_s^2$.
This relation allows to fix the value of $\alpha_2$, once $\alpha_1$ has been chosen through the 
relation 
\be
\frac{1-\alpha_1}{1-\alpha_2}=\sqrt{\frac{m_2}{m_1}}
\label{eq:alpha}
\ee
This way, a density ratio of $1:1000$ can be obtained,
as shown in figure \ref{fig.2}, where a droplet of fluid 
of density $\rho_2=1$ is in equilibrium with a surrounding second component 
of density $\rho_1=1/1000$.
Notably, this configuration has been obtained with $\alpha_1=0,\alpha_2=0.998$ 
with a strength of the interaction potential 
imposed to be $g_{12}=g_{21}=0.5$.
Unfortunately, this is valid in a static case,
 but it is not stable for dynamic ones, 
since the corresponding value of $c_{s}^2$ is so low that this
 component is affected by high-Mach
 number effects and, thus, cannot be correctly described in terms of our LB model.
Furthermore, we have carried out numerical simulations varying the density ratio and
we have found that  the highest density ratio attainable in our framework without 
the arising of instabilities is $1:3$.
This result is in agreement with what was found in a recent article devoted to this issue \cite{McC_04}, but it is the first time that a complete dynamical test-case is considered.
More specifically, the values chosen for this configuration are:
$\alpha_1 = 4/9,\alpha_2=0.8$ with the strength of interaction potential 
imposed to be $g_{12}=g_{21}=0.065$.
Recently, it has been pointed out that higher density ratios 
may be obtained also in microflows considering self- and cross-collisions~\cite{Arc_07}. 
Nevertheless, this strategy is based upon complex modelling and interpolation techniques,
which could result computationally too demanding for general dynamical cases.
Therefore, in order to model the usual situation of two fluids with very different densities as
air and water or water in equilibrium with its vapour (in these cases the density ratio is $\approx 500$), 
we propose to work on viscosity values.
Looking at eq. (\ref{eq:front1}), it is readily seen that the asymptotic behaviour is 
controlled just by the ratio $\mu_1/\mu_2$ and is independent of $\rho_1,\rho_2$.
Thus, keeping a density ratio of $3$, we can describe different binary fluids changing appropriately 
the value of kinematic viscosity in order to get the correct ratio between the dynamical viscosities.
 This is sufficient to obtain a correct macroscopic description of the flow.
In particular, we have chosen:
\be
\rho_1=0.34; \; \rho_2=1.; \; \nu_1 = 0.042 \;  \nu_2= 0.667
\label{eq:param}
\ee
and, as described above, 
\be
\alpha_1 = 4/9; \; \alpha_2=0.8; \;  g_{12}=g_{21}=0.065.
\ee
This choice corresponds to a dynamical viscosity ratio which can approximately
describe the case of the capillary filling of water in air.
In this configuration the surface tension has been computed to be $\gamma \approx 0.02$
via the Laplace's law which states that at equilibrium $\gamma=\frac{\Delta p}{R}$ 
for a drop of radius $R$.

\begin{figure}
\resizebox{1.0\columnwidth}{!}{%
  \includegraphics{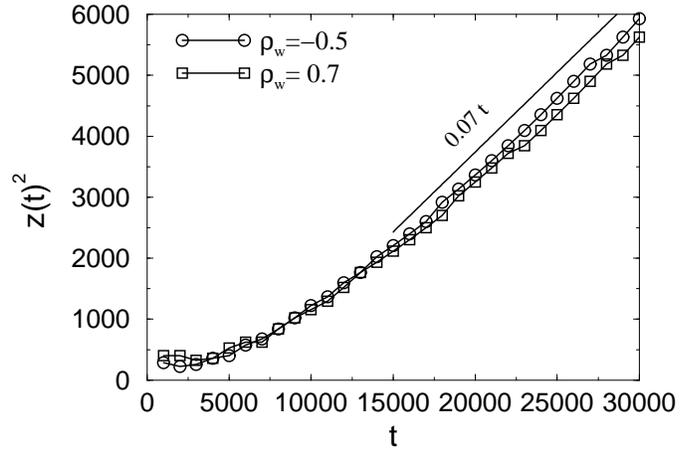}
}
\caption{Front displacement $z^2(t)$ with time for a given wettability, in this case it is $\theta \approx 35$ . The circle and square curves refer to the two different ways of imposing a given contact angle discussed in this work. First, it is possible to let interact the first component (the incoming fluid) with the walls via an attractive force, this case is represented by the curve labelled by $\rho_w=-0.5$, where the minus sign underlines the fact that the force is attractive. Second, the second component interacts with the walls through a repulsive force, this case is represented by the second curve labelled by $\rho_w=0.7$.
The straight solid line represents the asymptotic law, eq. (\ref{eq:wash}), with a contact angle $\theta=35$.}
\label{fig.4}
\end{figure}

We have seen that the desired wettability, that is a given contact angle, 
can be introduced in our model
through a phenomenological force~\cite{Kan_02,ben_06}, whose expression is given by (\ref{forcingw}).
Thus, the interaction with the walls is characterised by 
a phenomenological force proportional to $g_{1w} \rho_w$ and $g_{2w} \rho_w$, 
respectively for component $1$ and $2$, where $g_{1w}$ and $g_{2w}$ are independent.
Therefore, two different scenarios are possible for a given contact angle.
Referring to fig. \ref{fig.1}, we can let interact the incoming fluid $1$,
by imposing a non-zero value
of $g_{1w} \rho_{w}$ and giving to the force (\ref{forcingw}) the sign
to make the force attractive. 
On the other hand, it is also possible to impose a non-zero value
of $g_{2w} \rho_{w}$  and therefore to let interact the wall with the component $2$,
 but with the opposite sign,
such that in this case the fluid $2$ is repelled.
In both cases, the fluid $1$ feels hydrophilic  walls with respect to the other component.
However, in one case the fluid $1$ is directly pushed inside by the attraction of walls,
in the other it enters, since the walls are hydrophobic with the fluid $2$.
In fig. \ref{fig.4}, we show that for the same geometric and physical 
configuration and with an appropriate value of $g_{kw}$ and $\rho_w$ the same behaviour is found and therefore 
from a macroscopic point of view these two methods can be considered equivalent.
In particular, for this experiment we have used:
$g_{1w}=-0.065;~ g_{2w}=0;$ and $\rho_{w}=0.5;$ for the first case and 
$g_{1w}=0;~ g_{2w}=0.065; ~\rho_{w}=0.7;$ in the second.
This choice corresponds to a contact angle $\theta \approx 35$.
Data show a good agreement with the asymptotic law
given by the Washburn's law, eq. (\ref{eq:wash}),
computed with our physical parameters and 
this contact angle.

\begin{figure}
\resizebox{1.0\columnwidth}{!}{%
  \includegraphics{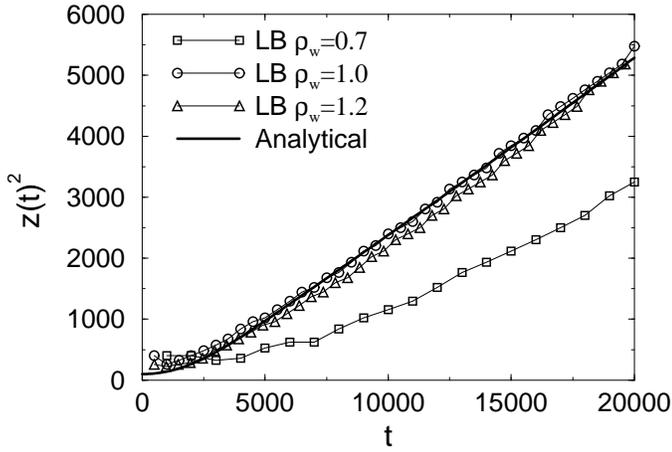}
}
\caption{Front displacement $z^2(t)$ with time for the case of complete wettability. Three numerical simulation profiles are displayed for three different values of the phenomenological constant $\rho_w$.
The curves show that the effect has reached his asymptotic value,
namely the complete wetting, for $\rho_w=1.0$.
The analytical solution for the case of complete wetting ($\theta=0$) is also shown. }
\label{fig.5}
\end{figure}

Then, we have concentrated our attention on the case of complete wetting. 
In order to obtain this configuration it is necessary to strengthen the interaction with the wall.
In figure \ref{fig.5}, we show results for the different configurations treated.
For the values of physical parameters described in eq. (\ref{eq:param}), 
 we found that the value of the contact angle increases with increasing  $\rho_{w}$ until the value of $\rho_{w}=1.0$. After this threshold is reached, all profiles collapse. 
In figure we show the curves corresponding to $\rho_{w}=0.7,~1.0$ and  $\rho_{w}=1.2$.
This behaviour indicates that the asymptotic value $\theta=0$ is reached.
Then,  we have computed the solution of the equation (\ref{eq:front1}) for the parameters chosen and $\theta=0$. The resulting solution is also shown in figure \ref{fig.5}. There is a good agreement between analytical and numerical solutions.

\begin{figure}
\resizebox{1.0\columnwidth}{!}{%
  \includegraphics{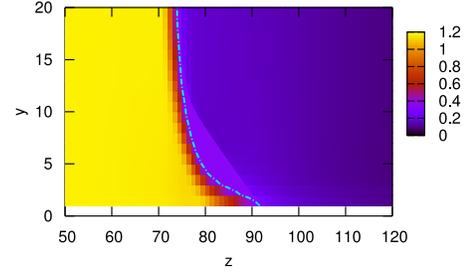}
}
\caption{ The figure shows the LB density profile in the vicinity of the meniscus at a given time ($t=20000$). Only the half of the channel is plotted ($y=20$) because of the symmetry. The picture represents the isocontours of density, whose values are given in the legend. The bulk density is about $1$, the density value far ahead  the interface tends to zero because of the immiscibility. The patterned colored line indicates the isocontour  $\rho=0.35$, which delimits the precursor film near the wall.}
\label{fig.6}
\end{figure}

\begin{figure}
\resizebox{1.0\columnwidth}{!}{%
  \includegraphics{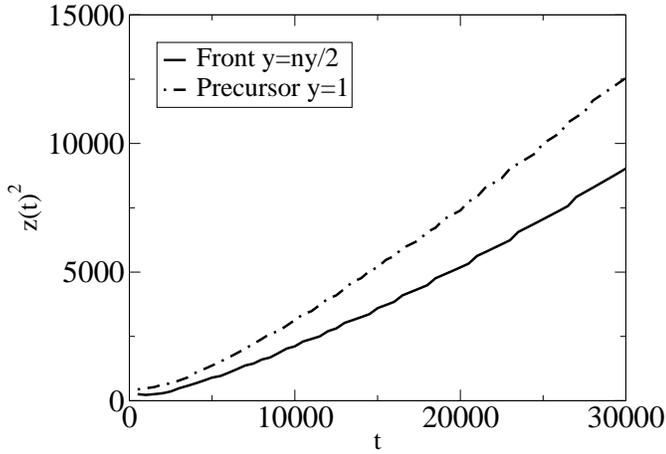}
}
\caption{ Front displacement $z^2(t)$ with time for the case of complete wettability as defined in fig. \ref{fig.4}. The dynamics of the interface is displayed  together with the position in time of the precursor, defined as the last point near-to-the-wall to have a density $\rho = \rho_{bulk}/3$. The two fronts advance in time with a different velocity.}
\label{fig.7}
\end{figure}

\begin{figure}
\resizebox{1.0\columnwidth}{!}{%
  \includegraphics{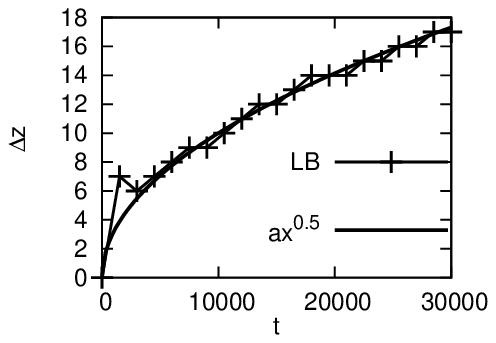}
}
\caption{The displacement of the precursor with respect to the central interface is plotted versus time. The curve is fitted with a square-root law $at^{1/2}$ with the coefficient that turns out to be $a=0.1$. This means that both fronts, the central and the precursor, advance in time with a square-root law but with a different pre-factor. The precursor  advances in time more rapidly.}
\label{fig.8}
\end{figure}

Finally, we have investigated the existence and the behaviour of precursor films,
whose presence is indicated by a thin layer of liquid near-to-the-wall which penetrates
the capillary ahead the bulk meniscus.
The presence and the importance of precursors film has been first recognised by De Gennes \cite{DeG_84}
and much attention has been devoted to it \cite{Bonn}, especially for the case of a spreading drop.
Some experiments have been attempted even at nanometric scale \cite{Kav_03}.
In our numerical simulations, we have observed that a thin layer ($1-2$ grid points) of liquid attached to the walls tends to penetrate faster, indicating the formation of a precursor film. 
The density of the film tends to decrease with the distance from the bulk of the liquid, since only some ``molecules'' escape from the bulk 
and they can not create a film as dense as the bulk liquid far away form the bulk interface.
We define the end of this film at the point where 
the density of the liquid becomes $1/3$ the bulk density.
In figure \ref{fig.6}, a snapshot of the density is shown which supports visually the presence 
of this precursor film. It is interesting to note that the layer 
is almost limited to the first point near to the wall, as observed in molecular dynamics simulations \cite{Dim_07}.
In figure \ref{fig.7}, we analyse the dynamical behaviour of precursors.
In figure \ref{fig.7}, the front displacement in time of both the front at the centre of the interface between the two components and the precursor (as defined above) are shown. 
It is clear that the precursor film  advances ahead the interface with a different law.
After $30000$ time-steps the precursor has attained the length of about 20 $\Delta x$.
This behaviour has been further analysed and the difference between the precursor position 
and the interface one is displayed in figure \ref{fig.8}. Both fronts advance in time 
according to a square-root law, but with different pre-factors. 
The displacement of the difference has been fitted
with a square-root with a pre-factor of $a=0.1$. 
In the very early stage ($t<5000$), the numerical curve of the difference shows a kink indicating that it advances faster than according to a square-root law, see fig. \ref{fig.8}.
This can be probably explained by the fact that,
during this initial time, the bulk front is affected 
by a ``vena contracta'' effect, which reflects the non-trivial matching between the reservoir 
and the capillary dynamics at the inlet~\cite{bif_07}.

These results appear to be in line with recent molecular dynamics findings \cite{Dim_07} and also with experiments at macroscopic scales~\cite{Bico}.

\section{Conclusions}

In this work, we have studied the penetration of one fluid into a capillary initially filled by another immiscible fluid by the means of a Lattice-Boltzmann model for binary fluids based upon a pseudo-potential interaction.

A phenomenological force has been introduced in order to let the interface to develop  a contact angle with the walls.
Two equivalent ways of imposing a given contact angle have been analysed.

In order to reproduce the classical experimental configuration, where  a much denser fluid penetrates into a much lighter fluid (vacuum or air), the model has been modified to let a density ratio between the two components and different kinematic viscosities have been used, such that the ratio between dynamical viscosities is comparable with that for a water-air mixture. Furthermore, it has been shown theoretically that this ratio is responsible for the dynamical behaviour of the filling. In this sense, we have studied two important asymptotic configurations:
one one hand, when the two components have equal density and viscosity, the front advances linearly in time. On the other hand, for a vanishing ratio $\mu_2/\mu_1$, the washburn's law is retrieved, which states that the front advances with a square-root law in time. 
Both asymptotic behaviours can be found out through the analytical solution of the equation of motion eq. (\ref{eq:front1}).

In a configuration of complete wetting ($\theta=0$) the displacement of the interface with time has 
been shown to agree with theoretical expectations.

Finally, the presence of a precursor film ahead the central front has been detected. 
This film occupies only a thin layer of one-two grid points and it moves with a square-root time law 
(like the interface at the centre of the channel), but with a different pre-factor, faster than the interface at the centre.
These results appear to be in line with recent molecular dynamics findings.

In this work, we present results for a chosen 2-D channel configuration.
Even though recent 3-D molecular dynamics results seem to suggest that the landscape 
is not much affected by the change of geometry, it will be important to simulate the imbibition
in different and more complex geometries in order to study the dependence of the phenomenon on
the simulation parameters. 
In particular, the behaviour of the precursor films could be more complex.


\section{acknowledgments}
S. Chibbaro's work is supported by a ERG EU grant.
This work makes use of results produced by the PI2S2 Project managed by the Consorzio COMETA, a project co-funded by the Italian Ministry of University and Research (MIUR).
He greatly acknowledges the financial support given also by the consortium SCIRE. 
More information is available at  http://www.consorzio-cometa.it. 
I would like to thank S. Succi for critical reading of this manuscript,
F. Diotallevi for the figure 1, 
and  L. Biferale and F. Toschi 
for helpful discussions.


\begin{thebibliography}{0}
\bibitem{DeG_84} P. G. de Gennes, Rev. Mod. Phys. {\bf 57}, 827 (1985)
\bibitem{tas} N.R. Tas et al., Appl. Phys. Lett. {\bf 85}, 3274,  (2004).
\bibitem{Wash_21} E.W. Washburn, Phys. Rev. {\bf 27} (1921) 273.
\bibitem{Lucas} R. Lucas, Kooloid-Z {\bf 23} (1918) 15.
\bibitem{Bos_23} C.H. Bosanguet Philos. Mag. Ser. 6 {\bf 45} 525 (1923).
\bibitem{Ske_71} J. Szekelely, A.W. Neumann, and Y.K. Chuang Journal of Coll. and Int. Science, {\bf 35} (1971) 273.
\bibitem{Mas_02} G. Mastic et al., Langmuir {\bf 18}, 7971,  (2002).
\bibitem{Sup_03} S. Supple and N. Quirke, Phys. Rev. Lett., {\bf 90}, 214501,  (2003).
\bibitem{Yan_04} L.-J. Yang, T.J. Yao, and Y.-C. Tai, J. Micromech. Microeng., {\bf 14}, 220,  (2004).
\bibitem{Jua_06} W. Juang, Q. Lui, and Y. Li, Chem. Eng. Technol., {\bf 29}, 716,  (2006).
\bibitem{frenkel} D. Rapaport, {\it The art of Molecular Dynamics Simulations} (Cambridge University Press, 1995).
\bibitem{Ben_92} R. Benzi, S. Succi, and Vergassola, The lattice-Boltzmann equation-theory and applications, Phys. Rep.  {\bf 222}, 145, (1992).
\bibitem{SS_92} R. Benzi, S. Succi, and Vergassola, The lattice-Boltzmann equation-theory and applications, Phys. Rep.  {\bf 222}, 145, (1992).
\bibitem{Che_98} S. Chen and G. Doolen, Annu. Rev. Fluid Mech. {\bf 30}, 329, 1998.
\bibitem{wolf}  D.A. Wolf-Gladrow  {\it Lattice-gas Cellular Automata and Lattice Boltzmann Models } (Springer, Berlin, 2000).
\bibitem{Squ_05} T. M. Squires and S. R. Quake, Rev. Mod. Phys. {\bf 77}, 977 2005. 
\bibitem{Tab_03} P. Tabeling, Introduction a la Microfluidique, Belin, Paris, 2003
\bibitem{Dup_05} A. Dupuis and J. M. Yeomans, Langmuir {\bf 21}, 2624 2005. 
\bibitem{Har_06} J. Harting, C. Kunert, and H. Herrmann, Europhys. Lett. {\bf 75}, 328 2006. 
\bibitem{Yar_06} A. L. Yarin, Annu. Rev. Fluid Mech. {\bf 38}, 159 2006. 
\bibitem{Rey_06} M. Reyssat, A. Pepin, F. Marty, Y. Chen, and D. Quèrè, Europhys. Lett. {\bf 74}, 306 2006.
\bibitem{Ric_02} D. Richard, C. Clanet, and D. Quèrè, Nature London {\bf 417}, 811 2002. 
\bibitem{SC_93} X. Shan, and H. Chen  Phys Rev E {\bf 47}, 1815, (1993). 
\bibitem{Sbr_06}  M. Sbragaglia,R. Benzi, L.  Biferale,  S. Succi,  and F. Toschi, Phys. Rev. Lett. {\bf 97} (2006) 204503.
\bibitem{Kwok_05} Zhang, J.; Kwok, D. Y., Langmuir  {\bf 20},  8137, (2004).
\bibitem{Kwok_06} Zhang, J.; Kwok, D. Y., Langmuir  {\bf 22},  4998, (2006).
\bibitem{Fab_07} F. Diotallevi, L. Biferale, S. Chibbaro, G. Pontrelli, F. Toschi, and S. Succi EpjB submitted.
\bibitem{Wol_04} L. Dos Santos, F. Wolf, and P. Philippi   J. Stat. Phys. {\bf 121}, 197 (2005).
\bibitem{Dim_07} D. I. Dimitrov, A. Milchev, and K. Binder, Phys. Rev. Lett. {\bf 99}, 054501 (2007)
\bibitem{Hub_07} P. Huber, K. Knorr, and A.V. Kityk, Mater. Res. Soc. Symp. Proc. 899E, N7.1 (2006).
\bibitem{Kan_02} Kang, Zhang and Chen Phys. Fluids {\bf 14} (9) 3203, (2002)
\bibitem{napoli} G. Cavaccini, V. Pianese, A. Jannelli, S. Iacono and R. Fazio, Lecture Series Comp. Comput. Sci. {\bf 7} (2006) 66.
\bibitem{Rot_07} M. Latva-Kokko, and D.H. Rothman   Phys Rev Lett {\bf 98}, 254503 (2007). 
\bibitem{Yeo_03} A. J. Briant and J. M. Yeomans Phys. Rev. E {\bf  69}, 031603 (2004)
\bibitem{Kar_06} S. Arcidiacono, J. Mantzaras, S. Ansumali, I. V. Karlin, C. Frouzakis, and K. B. Boulouchos Phys. Rev. E {\bf  74}, 056707 (2006)
\bibitem{McC_04} M.E. McCracken and J. Abraham  Phys. Rev. E {\bf 71}, 046704, (2005).
\bibitem{Arc_07} S. Arcidiacono,I.V. Karlin, J. Mantzaras, and C.E. Frouzakis  Phys. Rev. E {\bf 76}, 046703, (2007).
\bibitem{ben_06}  R. Benzi, L.  Biferale, M. Sbragaglia, S. Succi,  and F. Toschi, Phys. Rev. E {\bf 74} (2006)  021509.
\bibitem{Hua_07}  H. Huang, D. T. Thorne, M. G. Schaap, and M. C. Sukop, Phys. Rev. E {\bf 76}, 066701 (2007).
\bibitem{Bonn} D. Bonn, J. Eggers, J. Indekeu, J. Meunier, E. Rolley, submitted to  Rev. Mod. Phys.
\bibitem{Kav_03} H. Pirouz Kavehpour, Ben Ovryn, and Gareth H. McKinley Phys. Rev. Lett. {\bf 91}, 196104 (2003)
\bibitem{bif_07} F. Diotallevi, L. Biferale, S. Chibbaro, A. Lamura, G. Pontrelli, M. Sbragaglia, F. Toschi, and S. Succi EpjB submitted.
\bibitem{Bico} J. Bico, and D. Qu\'er\'e Europhys. Lett. {\bf 61}, 348 (2003).


\end{thebibliography}
\end{document}